\documentclass[manuscript, screen, nonacm]{acmart}
\usepackage{booktabs} 

\usepackage{cleveref}

\usepackage{lipsum}
\usepackage{amsfonts}
\usepackage{graphicx}
\usepackage{epstopdf}

\usepackage{algorithmicx}
\usepackage{algorithm}
\usepackage[noend]{algpseudocode}
\usepackage{cuted}
\usepackage{tikz}
\usepackage{tkz-euclide}
\usetikzlibrary{patterns}
\usetikzlibrary[patterns]
\usepgflibrary{patterns}
\usepgflibrary[patterns]

\begin{document}

\title{Online Total Completion Time Scheduling on Parallel Identical Machines}

\author{Uwe Schwiegelshohn}
\affiliation{%
  \institution{TU Dortmund}
  \country{Germany}
}
\email{uwe.schwiegelshohn@tu-dortmund.de}

\renewcommand\shortauthors{Schwiegelshohn, U.}

\begin{abstract}
We investigate deterministic non-preemptive online scheduling with delayed commitment  for total completion time minimization on parallel identical machines. In this problem, jobs arrive one-by-one and their processing times are revealed upon arrival. An online algorithm can assign a job to a machine at any time after its arrival. We neither allow preemption nor a restart of the job, that is, once started, the job occupies the assigned machine until its completion. Our objective is the minimization of the sum of the completion times of all jobs. In the more general weighted version of the problem, we multiply the completion time of a job by the individual weight of the job. We apply competitive analysis to evaluate our algorithms. 

We improve 25-year-old lower bounds for the competitive ratio of this problem by optimizing a simple job pattern. These lower bounds decrease with growing numbers of machines. Based on the job pattern, we develop an online algorithm which is an extension of the delayed-SPT (Shortest Processing Time first) approach to the parallel machine environment. We show that the competitive ratio is at most 1.546 for any even machine number which is a significant improvement over the best previously known competitive ratio of 1.791. For the two-machine environment, this algorithm achieves a tight competitive ratio. This is the first algorithm which optimally solves an online total completion time problem in a parallel machine environment. Finally, we give the first separation between the weighted and unweighted versions of the problem by showing that in the two-machine environment, the competitive ratio of the weighted completion time objective is strictly larger than 1.546. 
\end{abstract}

%
%

\ccsdesc[500]{Theory of computation~Scheduling algorithms}
\ccsdesc[500]{Theory of computation~Online algorithms}
%
%

\keywords{Online Algorithms, Scheduling and Resource Allocation, Total Completion Time}

\maketitle

\section{Introduction}
\label{sec:introduction}

Total completion time with or without weights is - apart from makespan - the most prominent objective in scheduling problems. We look for a schedule that minimizes the sum of the (weighted) completion times of the jobs. A job $J_j$ is characterized by its processing time $p_j$ and possibly its weight $w_j$. The most basic problems use either a single machine or parallel identical machines as their environment. In online scheduling problems, jobs arrive one-by-one such that the properties of job $J_j$ are revealed at its submission time $r_j$. Competitive analysis is the most common approach to evaluate those problems, that is, we determine the ratio of an online result over the corresponding offline result and consider the worst case. Here, we address non-preemptive online scheduling with delayed commitment in a parallel identical machine environment and primarily assume no weight for the jobs. More precisely, we can assign a job to any available machine at any time after its submission. Once started, the job occupies the assigned machine until it completes, that is, we do not allow preemption or a restart of the job. 

The corresponding makespan problem is well studied. Typically, it requires an immediate commitment although, more recently, \cite{EOW14} used the comparable reorder buffer approach which is related to delayed commitment. Graham's simple list scheduling algorithm \cite{Gra66} achieves an optimal competitive ratio $2-\frac{1}{m}$ for $m=2$ and $m=3$ machines. However, even after many research efforts, there is still a gap between the competitive ratio of the best online algorithm and the lower bound of the competitive ratio for the makespan in the parallel machine environment if the number of machines is at least 4, see \cite{FKT89}, although the gap is very small (0.0665)~\cite{FlW00}.    

The online total (weighted) completion time problem on a single machine is solved since there are algorithms which match the lower bound $2$ of the competitive ratio, see, for instance, \cite{Hoogeveen1996}. One of these algorithms is  
the delayed-SPT (Shortest Processing Time first) approach which maintains an SPT-ordered list of all jobs that are submitted but not yet started, that is, we update the SPT-list whenever a job is started or a job is submitted. Then the delayed-SPT algorithm starts the first job at the head of the list as soon as the machine becomes available provided some absolute delay has passed since the beginning of the schedule at time 0. This delay only depends on the processing time of the job. Since the algorithm only considers the first job of the SPT-list, it works in a local fashion. 

However, for the parallel identical machine environment, there are still some questions that are open for a long time. In particular, there is a significant gap ($\geq 0.271$) between the best known lower~\cite{Vestjens1997} and upper bounds~\cite{Sit10} of the competitive ratio. Also we do not know whether weights result in a problem separation, that is, whether the lower bound of the competitive  ratio for a problem with weights is strictly larger than the best competitive ratio for the corresponding problem without weights.

\subsection{Our Contribution}
\label{sec:contribution}

In this paper, we improve the existing lower bounds of \cite{Vestjens1997} by optimizing the used job pattern. This optimization leads to a system of equations that require numerical solutions. 

Then we introduce a simple online algorithm which is an extension of the delayed-SPT approach. We show that the algorithm guarantees the competitive ratio $1.546$ for all environments with an even number of machines. This competitive ratio is significantly better than the best previously known competitive ratio of $1.791$~\cite{Sit10}. Moreover, it is tight for the two-machine environment. This is the first time that an online algorithm achieves an optimal competitive ratio for the total completion time problem in a parallel environment. The analysis is tailored to the two-machine environment. To show further improvements for other environments, we need a slightly different set-up. However, the delayed-SPT algorithm that only considers the first job of the SPT-list cannot achieve the lower bound of the competitive ratio for environments with more than 2 machines.   

Further, we establish a separation between the total completion time objectives with and without weights by showing that in the two-machine environment, the total weighted completion time problem has a strictly larger competitive ratio than $1.546$. 

\subsection{Related Work}
\label{sec:related-work}

Minimizing the total completion time is one of the most fundamental scheduling problem.  Without release dates, we can solve the problem in the single machine and the parallel identical machine environments by using the Shortest Processing Time first (SPT) rule~\cite{CMM67}.    

In presence of release dates, the problem on a single machine is already strongly NP-hard~\cite{Lenstra1977}. It becomes easy if we allow preemption. Then we can use the preemptive version of SPT: the Shortest Remaining Processing Time first rule (SRPT)~\cite{Lawler1993}. In the parallel identical machine environment, the problem with preemption is also NP-hard~\cite{Du1990}. The preemptive problem with weights on a single machine is already strongly NP-hard~\cite{LLLR84}. Since the discovery of those basic results, there has been ongoing research activity in this area, see, for instance, \cite{Li17,BSS16}.

For the online scheduling problem on a single machine, several authors - apart from \cite{Hoogeveen1996} - have presented optimal online algorithms: Phillips, Stein, and Wein~\cite{Phillips1998} proposed an algorithm based on the optimal preemptive schedule. Stougie (cited in Vestjens~\cite{Vestjens1997}) came up with a third algorithm using again shifted release dates. Other 2-competitive algorithms have been discovered subsequently~\cite{Lu2003,Goemans1997}

On $m$ parallel identical machines,  Chekuri, Motwani, Nataranjan, and Stein~\cite{Chekuri2001} suggested a $3-\frac{1}{m}$-competitive algorithm. First they create a preemptive schedule on one machine and then they develop a non-preemptive schedule on identical machines based on the order of completion times of the job in a one-machine-relaxation schedule. \cite{MeS04} introduced the algorithm Shifted WSPT and improved the result to a competitive ratio of $2+\max \{1/\alpha, \alpha + (m-1)/2m\}$. \cite{LiL09} provided another improvement to 2. \cite{Sit10} addressed the weighted completion time problem with preemption on a single machine and used this result to obtain online schedules for parallel identical machines. This algorithm achieves the currently best known competitive ratio 1.791. This is the first result beating the single machine competitive ratio of 2. Note that Sitters' result holds for randomized and deterministic scheduling with and without weights. Already in 1997, \cite{Vestjens1997} has presented lower bounds of the competitive ratio for the problem. The values of these lower bounds depend on the number of machines and range from 1.520 for $m=2$ to 1.309 for infinitely many machines. For the related preemptive online problem, \cite{CNS10} showed that SRPT has a better competitive ratio than 2. Shortly afterwards, \cite{Sit10} improved this  result to 1.25.    

There are also related results. For instance, \cite{AvA07} do not use the delayed commitment model but require an immediate selection of a machine for each job. They compare their online algorithm with an optimal offline algorithm using migration and mainly consider the total completion time and the total flow time objectives. L{\"u}bbecke, Maurer, Megow, and Wiese apply a different approach~\cite{LMMW16}. They approximate the optimal competitive ratio and use the total (weighted) completion time problem as an application area.  

\subsection{Outline}
\label{sec:outline}

After describing some additional notations, we explain a key pattern that forms the foundation of the lower bounds and of our algorithm. This pattern leads to an optimization problem that is specific for the machine environment. We translate this optimization result into a lower bound for the competitive ratio of our online problem. Afterwards we introduce the concept of our online algorithm and describe it formally. Then we prove the competitive ratio of the algorithm for $m=2$ and extend it  to environments with an even number of machines. Finally, we show that a lower bound of the competitive ratio for the weighted problem in the two-machine environment is strictly larger than the competitive ratio of our algorithm for the problem without weights in the same environment.  

\section{Notations}
\label{sec:notations}

The jobs of our problem form a sequence ordered by increasing submission times such that $r_1=0$.  Variables $C_j(S)$ and $C_j^*$ denote the completion times of job $J_j$ in schedule $S$ and in an optimal schedule, respectively. We also use $C_{max}(S)$ and $C_{max}^*$ for the makespan in schedule $S$ and an optimal total completion time schedule, respectively. If there is no chance of confusing two schedules then we omit the dependence on $S$. In an SPT-schedule, job $J_k$ does not start before job $J_j$ for any pair of jobs $J_j$ and $J_k$ with $p_j<p_k$. If a job $J_j$ has the next start time after a job $J_k$ with $p_j<p_k$ then we say that job $J_j$ establishes an \textit{inversion}. A non-delay SPT-schedule optimally solves the total completion time problem if all jobs are available at time 0.

\section{Key Pattern}
\label{sec:pattern}

We introduce a simple key pattern comprising $k$ identical jobs with processing time $\delta$ and submission time $r$. In the \textit{restricted} schedule $S_r$, machine $i\in\{ 1,\ldots, m\}$ becomes available for job processing at time $m_i\geq r$. We are interested in the ratio $\mathcal{R}$ between the total completion time $\sum C(S_r)$ of these jobs in schedule $S_r$ and their optimal total completion time $\sum C^*$ in an optimal schedule where all machines are available at time $r$.

If machine $i$ executes $k_i$ identical jobs in a non-delay schedule $S_r$ then the total completion time of these jobs is 
\begin{align}
\label{eq:single_comp}
k_i\cdot m_i + \frac{k_i\cdot (k_i+1)}{2}\cdot \delta  &= k_i \cdot 
(m_i + \frac{1}{2}\cdot k_i\cdot \delta +\frac{1}{2}\cdot \delta) = 
\frac{(m_i+k_i\cdot \delta)^2-m_i^2}{2\delta} + \frac{k_i\cdot \delta}{2},
\end{align}
resulting in the total completion time
\begin{align}
\label{eq:total_comp}
\sum C(S_r) &= \frac{\sum_{i=1}^m \left((m_i+k_i\cdot \delta)^2-m_i^2\right)}{2\delta} + \frac{k\cdot \delta}{2}.
\end{align}
for schedule $S_r$. Due to the non-delay property, the makespans of machines $i$ and machine $h$ differ by at most processing time $\delta$ if both machines execute at least one job, that is, for $\delta\rightarrow 0$, two machines in a non-delay schedule have the same makespan unless one machine becomes available after the makespan. We assume an arbitrarily small processing time $\delta$. For a given total processing time, this assumption results in a large number of jobs and we can ignore the contribution of all other jobs to the total completion time. Then Eq.~\eqref{eq:single_comp} and \eqref{eq:total_comp} yield 
\begin{align}
\label{eq:delta_ratio}
\lim_{\delta\rightarrow 0} \mathcal{R} & = \lim_{\delta\rightarrow 0}\frac{\sum C(S_r)}{\sum C^*} = \lim_{\delta\rightarrow 0} \frac{\sum_{i=1}^m \left((m_i+k_i\cdot \delta)^2-m_i^2\right)+k\cdot \delta^2}{\sum_{i=1}^m \left((r+k_i^*\cdot \delta)^2-r^2\right)+k\cdot \delta^2}  = \frac{\sum_{i=1}^m \left(\left(\max\{m_i,C_{max}(S_r)\}\right)^2-m_i^2\right)}{m\cdot \left(\left(C_{max}^*\right)^2 -r^2\right)}.
\end{align}
In Eq.~\eqref{eq:delta_ratio}, $k_i^*$ denotes the number of jobs executed by machine $i$ in the optimal schedule.

We limit any value $m_i$ to $C_{max}(S_r)$ since such limit does not change the total completion time in schedule $S_r$. Then we have $k \cdot \delta = m\cdot C_{max}(S_r) - \sum_{i=1}^m m_i$ and obtain 
\begin{align}
\label{eq:calc_ratio}
\mathcal{R} & = \frac{\sum_{i=1}^m \left(\left(C_{max}(S_r)\right)^2-m_i^2\right)}{m\cdot \left( \left(C_{max}(S_r)+ r -\frac{\sum_{i=1}^m m_i}{m} \right)^2-r^2\right)} = \frac{  \left(C_{max}(S_r)\right)^2-\frac{\sum_{i=1}^m m_i^2}{m}}{\left( C_{max}(S_r) -\frac{\sum_{i=1}^m m_i}{m} \right)^2+ 2r \cdot \left( C_{max}(S_r) -\frac{\sum_{i=1}^m m_i}{m} \right)}. 
\end{align}
In addition, we define the \textit{makespan ratio} $ \mathcal{R}_{max}$:    
\begin{align}
\label{eq:calc_limit}
\mathcal{R}_{max} & =  \frac{C_{max}(S_r)}{C_{max}^*} = \frac{C_{max}(S_r)}{C_{max}(S_r)-\frac{\sum_{i=1}^m m_i}{m}+r} 
\end{align}
We are interested in the makespan $C_{max}^{max}(S_r)$ that produces the largest ratio for a given set of values $m_i\leq C_{max}^{max}(S_r)$. Algorithm~\ref{alg:delay} determines this value. It starts with the makespan $C=\max_{1\leq i\leq m}\{m_i\}$ and the previous delay value $r$ or $0$ if there is no such previous delay value. For $\mathcal{R}_{max}(C)<\mathcal{R}(C)$, we know $C_{max}^{max}(S_r)<C$ and use the next smaller machine availability value as makespan $C$. To this end, we reduce all larger machine availability values in Eq.~\eqref{eq:calc_ratio} to $C$.

Assume $\mathcal{R}_{max}(\max\{ m_i\})\geq \mathcal{R}(\max\{ m_i\})$. Then we use Eq.~\eqref{eq:calc_ratio} and \eqref{eq:calc_limit} to determine the dependence of $r$ from the target ratio $\mathcal{R}_m$ by eliminating the makespan $C_{max}^{max}(S_r)$ since $\mathcal{R}_{max}(C_{max}^{max}(S_r))=\mathcal{R}(C_{max}^{max}(S_r))=\mathcal{R}_m$ holds. We obtain the equation
\begin{align}
\label{eq:calc_release}
(r\cdot \mathcal{R}_m)^2 -2r\cdot \mathcal{R}_m\cdot \frac{\sum_{i=1}^m{m_i}}{m}+ \mathcal{R}_m\cdot \left(\frac{\sum_{i=1}^m{m_i}}{m} \right)^2 -(\mathcal{R}_m-1)\cdot \frac{\sum_{i=1}^m{m_i^2}}{m} & =0 \;\;\; \mbox{ and its solution } 
\end{align}
\begin{align}
\label{eq:calc_solution}
r & = \frac{1}{\mathcal{R}_m} \cdot \left(\frac{\sum_{i=1}^m{m_i}}{m} - \sqrt{(\mathcal{R}_m-1)\cdot \left(\frac{\sum_{i=1}^m{m_i^2}}{m}-\left(\frac{\sum_{i=1}^m{m_i}}{m} \right)^2  \right)}  \right).
\end{align}
We only use the minus-operator in front of the square root in Eq.~\eqref{eq:calc_solution} since the submission time $r$ of the small jobs decreases with an increasing value of $\frac{\sum_{i=1}^m{m_i^2}}{m}$ if $\mathcal{R}_m$ and $\frac{\sum_{i=1}^m{m_i}}{m}$ are constant.

Alternatively, we obtain Eq.~\eqref{eq:calc_release} by determining the maximum of $C_{max}(S_r)$ in Eq.~\eqref{eq:calc_ratio}. If we know $r$ then Eq.~\eqref{eq:calc_release} produces ratio $\mathcal{R}_m$. To calculate the submission time $r$ of the jobs, we must guarantee $m_i\geq r$ for all machines, see Line~9 of Algorithm~\ref{alg:delay}. Then we obtain a quadratic equation for the variable $r$. The solution is valid if $m_i\geq r$ holds for all machines. Otherwise, we must replace violating values $m_i$ by $r$ and repeat these steps, see Lines~8 to 11 of Algorithm~\ref{alg:delay}.  

The next lemma expresses the dependence between the submission time $r$ of the small jobs and the ratio $\mathcal{R}$.
\begin{lemma}
\label{lem:release_date}
Any increase of the common submission time $r$ of the small jobs decreases the ratio $\mathcal{R}$ of Eq.~\eqref{eq:calc_ratio}. 
\end{lemma}

\begin{proof}
An increase $\delta_r$ of $r$ increases the contribution of each small job to the total completion time $\sum C_j^*$ of the optimal schedule by $\delta_r$ while it increases the contribution of a small job to the total completion time $\sum C_j(S_r)$ of the restricted SPT schedule $S_r$ by at most $\delta_r$ since the increase of the submission time may not or only partially affect the machine availability.
\end{proof}

Informally, the lemma states that we must concentrate on the first time intervals relative to the processing time for our analysis since later time intervals have less influence on the ratio $\mathcal{R}$. In Section~\ref{sec:two_machines}, Eq.~\eqref{eq:small_alpha} and \eqref{eq:large_alpha} describe this relation in a more formal way for the two-machine environment.

\begin{algorithm}[t]
\caption{Delay Detection}
\label{alg:delay}
\begin{algorithmic}[1]
\State {$C=max\{m_i\}$}
\State {$r=max\{r_j, \mbox{ previous delay value }\}$}
\While {$\mathcal{R}_{max}(C)< \mathcal{R}(C)$ using Eq.~\eqref{eq:calc_ratio} and \eqref{eq:calc_limit}}
\State {reduce $C$ to the next smaller $m_i$ value}
\State {reduce all $m_i>C$ in Eq.~\eqref{eq:calc_ratio} and \eqref{eq:calc_limit} to $C$}
\EndWhile

\State {determine ratio $\mathcal{R}$ using $r$ in Eq.~\eqref{eq:calc_release}}
\If {$\mathcal{R} >$ target ratio $\mathcal{R}_m$}
\Repeat
\State {replace all $m_i<r$ with $r$}
\State {determine the new delay value $r$ using Eq.~\eqref{eq:calc_release} with $\mathcal{R}_m$}
\Until {$m_i\geq r$ holds for all $m_i$ values}
\EndIf 
\end{algorithmic}
\end{algorithm}

\section{Lower Bounds}
\label{sec:lower_bounds}

Since we do not know the target ratio $\mathcal{R}_m$ in Eq.~\eqref{eq:calc_solution}, we obtain it by defining and solving an optimization problem. Informally, we consider an algorithm that uses some form of delayed-SPT with different delay values. To determine these values, we assume $m$ identical \textit{long} jobs $J_j$ with processing time $p_j=1$ and submission time $r_j=0$. We start the $i$th long job on machine $i$ with delay $t_i$. Then machine $i$ becomes available at time $t_i+1$. Immediately after the start of the long job on machine $i$, there may be a submission of the small job pattern of Section~\ref{sec:pattern} with the common submission time $r=t_i$. Note that the remaining $m-i$ machines are available for the small jobs at their submission time $r$ since the schedule has only started $i$ long jobs so far. 

Starting with some target ratio $\mathcal{R}$, Algorithm~\ref{alg:delay} uses Eq.~\eqref{eq:calc_solution} to determine delay $r=t_i$ in iteration $i$ and afterwards updates machine availability $m_i=t_i+1$. To verify our assumption of the target ratio $\mathcal{R}$, we consider the scenario with $m$ long jobs and without any small jobs. Our assumption is correct if the ratio $\mathcal{R}_c$ between the total completion time of the online schedule with $m$ delayed jobs and the corresponding optimal total completion time matches $\mathcal{R}$: 
\begin{align}
\label{eq:only_long_ratio}
\mathcal{R}_c & = \frac{m + \sum_{i=1}^m t_i}{m} = 1 + \frac{\sum_{i=1}^m t_i}{m}=\mathcal{R}
\end{align}
If $\mathcal{R}_c$ of Eq.~\eqref{eq:only_long_ratio} exceeds $\mathcal{R}$ then we increase $\mathcal{R}$ for the next run of the algorithm. Conversely, we decrease $\mathcal{R}$ for $\mathcal{R}_c<\mathcal{R}$. Note that the detection of the optimum target ratio precedes the online algorithm and is not part of it. A logarithmic search increases the efficiency of the algorithm.  

The next Lemma~\ref{lem:unique_delay} shows that for each number of machines $m$, there is a single set of time delays that achieves the correct target ratio $\mathcal{R}_m$.
\begin{lemma}
\label{lem:unique_delay}
There is a unique set of time delays that produces the minimum target ratio $\mathcal{R}_m$ for a given number $m$ of machines.
\end{lemma} 

\begin{proof}
Due to Eq.~\eqref{eq:only_long_ratio}, any change of the time delays must include the increase of some time delays and we must compensate an increase of time delay $t_i$ with $i\in \{1, \ldots, m\}$ by decreasing other time delays. Without restriction of generality, let $i$ be the index of the first machine with an increased time delay. 

Due to Lemma~\ref{lem:release_date}, the decrease of the positive time delay on any machine preceding machine $i$ violates the target ratio. Let machine $k$ with $k>i$ be the first machine after machine $i$ that starts its long job earlier than the corresponding delay $t_k$. Since the increase of $t_i$ results in the same increase of the machine availability of machine $i$, any decrease of $t_k$ produces a violation of the target ratio, see Eq.~\eqref{eq:calc_ratio}, and an increase of $t_i$ is not compatible with $\mathcal{R}_m$.
\end{proof} 

\cite{Vestjens1997} used a similar optimization problem to determine the previous set of lower bound values. Like \cite{Vestjens1997}, we have not found an explicit description of the solution for our optimization problem even for $m=2$. Therefore, we have numerically determined the optimal ratio $\mathcal{R}_m$ for several values of $m$ separately. We show some of the ratio results in Table~\ref{tab:ratio}, all of the corresponding delay results for some machine numbers in Table~\ref{tab:ti}, and the largest delay $t_m$ for further examples in Table~\ref{tab:tm}. 

\begin{table}[ht]
\begin{center}
\begin{tabular}{|c|c|c|c|c|c|c|c|c|c|c|}
\hline
$m$ & 2 & 3 & 4 & 5 & 10 & 100 & 1000 & 10000 & 100000 & 1000000 \\ \hline 
$\mathcal{R}_m$ & 1.54610 & 1.45961 & 1.44219 & 1.42907 & 1.39529 & 1.36905 & 1.36676 & 1.36651 &  1.36648 & 1.36648 \\
\hline
\end{tabular}
\end{center}
\caption{\label{tab:ratio} Target ratio $\mathcal{R}_m$ for different numbers $m$ of machines}
\end{table}

\begin{table}[ht]
\begin{center}
\begin{tabular}{|c|c|c|c|c|c|c|c|c|c|}
\hline
$m$ & $2$ & $3$ & $4$ & $5$ & $6$ & $7$ & $8$ & $9$ & $10$   \\ \hline
$t_1$ & 0.23898 & 0.02991 & 0 & 0 & 0 & 0 & 0 & 0 & 0     \\
$t_2$ & 0.85321 & 0.50413 & 0.30180 & 0.16459 & 0.06718 & 0 & 0 & 0 & 0   \\
$t_3$ & $--$ & 0.84479 & 0.61584 & 0.44983 & 0.32491 & 0.22774 & 0.14751 & 0.08259 &  0.02397  \\
$t_4$ & $--$ & $-- $ & 0.85185 & 0.67546 & 0.53758 & 0.42718 & 0.33596 & 0.25969 & 0.19537  \\
$t_5$ & $--$ & $--$ & $--$ & 0.85549 & 0.71155 & 0.59435 & 0.49730 & 0.41439 & 0.34308 \\
$t_6$ & $--$ & $--$ & $--$ & $--$ & 0.85607 & 0.73528 & 0.63520 & 0.54851 & 0.47295 \\
$t_7$ & $--$ & $--$ & $--$ & $--$ & $--$ & 0.85585 & 0.75411 & 0.66526 &  0.58711  \\
$t_8$ & $--$ & $--$ & $--$ & $--$ & $--$ & $--$ & 0.85803 & 0.76780 & 0.68805  \\
$t_9$ & $--$ & $--$ & $--$ & $--$ & $--$ & $--$ & $--$ & 0.85888  & 0.77804    \\
$t_{10}$ & $--$ & $--$ & $--$ & $--$ & $--$ & $--$ & $--$ & $--$ & 0.85895 \\
\hline
\end{tabular}
\end{center}
\caption{\label{tab:ti} Delays $t_i$ for different numbers $m$ of machines}
\end{table} 

\begin{table}[ht]
\begin{center}
\begin{tabular}{|c|c|c|c|}
\hline
$m$ & 10 & 100 & 1000 \\ \hline
$t_m$ & 0.85895 & 0.86414 & 0.85999 \\
\hline
\end{tabular}
\end{center}
\caption{\label{tab:tm} Delay $t_m$ produced by the optimization problem for different numbers $m$ of machines}
\end{table} 

Since \cite{Vestjens1997} uses the same scenario of identical long jobs in combination with many small jobs, the ratio $\mathcal{R}_m$ cannot be smaller than the previous lower bound for $m$ machines, due to Lemma~\ref{lem:unique_delay}. Finally, we formally show that the target ratio $\mathcal{R}_m$ is a lower bound for the competitive ratio in an $m$-machine environment.

\begin{theorem}
\label{thm:lbu}
Any deterministic online algorithm has at least competitive ratio $\mathcal{R}_m$ for the minimization of the total completion time with delayed commitment on $m$ parallel identical machines. 
\end{theorem}

\begin{proof}
The adversary submits $m$ identical long jobs $J_1$ with processing time $p_1=1$ and release date $r_1=0$. Due to Lemma~\ref{lem:unique_delay}, $ALG$ cannot compensate a larger time delay than the optimal delay for any long job. Therefore, if an online  algorithm $ALG$ selects a larger time delay than the optimal time delay for any long job, Eq.\eqref{eq:only_long_ratio} results in a violation of $\mathcal{R}_m$. If an online algorithm $ALG$ selects for any long job a smaller time delay than the optimal time delay then the adversary uses the pattern of Section~\ref{sec:pattern}. The resulting ratio of the total completion time of small jobs in the online schedule over the total completion time of these jobs in the optimal schedule violates $\mathcal{R}_m$. Since the arbitrarily small processing time of the small jobs results in an arbitrarily large number of small jobs, we can neglect the contribution of the long jobs in the online schedule and the optimal schedule.
\end{proof}

The rest of this section discusses properties of the solutions. Since the delays and the target ratio depend on the number of machines and we have no explicit formula for the solution, we give some upper bounds for these values. 
\begin{lemma}
\label{lem:opt_delay}
For $m>1$, we have $t_i < 1$ for $i\in\{1,2,\ldots ,m\}$ and $\mathcal{R}_m<2$. 
\end{lemma}

\begin{proof}
We use a simple induction approach in the machine index and assume $t_i=1$ and $t_h<1$ for $h\in\{1,2,\ldots, i-1\}$. Then Eq.~\eqref{eq:calc_ratio} produces a target ratio of less than 2 since all previous machines and all following machines are available before 2 and at 1, respectively. For $i<m$, we reduce $t_i$ such that we still have a target ratio less than two, see Lemma~\ref{lem:release_date}.

For $i=m$, the ratio of Eq.~\eqref{eq:calc_limit} is less than 2 since all other delays are less than 1. Therefore, we can also reduce $t_m$ and obtain a target ratio of less than 2.
\end{proof} 

Next, we address the smallest delay of a long job: for all scenarios with $m>3$, at least one machine starts the assigned long job without any delay, see Table~\ref{tab:ti}, since $r=0$ does not produce the target value $\mathcal{R}_m$. To determine how many machines start their jobs without delay, we use a continuous extension of the expressions in Eq.~\eqref{eq:calc_ratio} and \eqref{eq:calc_limit} for target ratio $\mathcal{R}_m$ 
\begin{align*}
\mathcal{R}_{max} =  \frac{C_{max}^{max}(S)}{C_{max}^{max}(S)-\frac{k}{m}} & = \frac{(C_{max}^{max}(S))^2-\frac{k}{m}}{\left( C_{max}^{max}(S) -\frac{k}{m} \right)^2} = \mathcal{R}_m 
\end{align*}
and obtain the optimal makespan $C_{max}^{max}(S)=1$ and the value $\frac{k}{m}= 1 - \frac{1}{\mathcal{R}_m}$ at which the delay becomes positive. Therefore, the first $\lfloor k \rfloor$ machines start their jobs at time 0 in our scenario.

The following lemma supports the observation that $\mathcal{R}_m$ decreases with increasing $m$, see Table~\ref{tab:ratio}.
\begin{lemma}
\label{lem:trend}
For positive integers $m$, we have $\mathcal{R}_{2m} < \mathcal{R}_m$. 
\end{lemma}

\begin{proof}
In a system with $m$ machines, we split every machine into two descendant machines. In particular, we consider machine $i$ with $t_i>0$ of the original system. At first, the long jobs on both of its descendant machines have the same delay value $t_i$. Therefore, the ratio of Eq.~\eqref{eq:only_long_ratio} remains unchanged.

If we apply our key pattern to the first descendant machine then the availability of this machine remains unchanged at time $t_i+1$ while the second descendant machine is available at time $t_i$. Therefore, the resulting ratio of Eq.~\eqref{eq:calc_ratio} is less than $\mathcal{R}_m$ and we reduce the delay of the job allocated to this first descendant machine. Due to this reduction, the first descendant machine becomes available before $t_i+1$ when considering the second descendant machine. Again the ratio of Eq.~\eqref{eq:calc_ratio} is less than $\mathcal{R}_m$ and we also reduce the delay of the job allocated to this machine. Due to these delay reductions, the ratio of Eq.~\eqref{eq:calc_limit} is less than $\mathcal{R}_m$ as well. Therefore, the minimal ratio $\mathcal{R}_{2m}$ is less than $\mathcal{R}_m$.
\end{proof}

We use a similar approach as in the proof of Lemma~\ref{lem:trend} to extend the performance evaluation of a two-machine environment to an environment with $2m$ machines, see Theorem~\ref{thm:even_number} in Section~\ref{sec:more_machines}.

\section{Algorithm}
\label{sec:algorithm}

\begin{algorithm}[t]
\caption{Online Total Completion Time Scheduling}
\label{alg:online}
\begin{algorithmic}[1]
\Repeat
\State {maintain a SPT order of all unscheduled jobs}
\If {the first position of the list has changed}
\State {determine the delay of the first job using Algorithm~\ref{alg:delay} with target ratio $\mathcal{R}_m$}
\EndIf
\If {a machine is available}
\If {the delay of the first job of the list is reached or in the past}
\State {start the first job of the list on the available machine}
\EndIf
\EndIf 
\Until {all jobs are started}
\end{algorithmic}
\end{algorithm}
We use the pattern of Section~\ref{sec:pattern} and Algorithm~\ref{alg:delay} in online Algorithm~\ref{alg:online}. Algorithm~\ref{alg:online} maintains an SPT-order of all unscheduled jobs by inserting any newly submitted job into the SPT-list. Afterwards, it determines the delay for the first job in this list based on the schedule generated so far. A new delay calculation is only necessary if another job has replaced this job at the head of the list. This replacement also requires a new calculation for the previous job once it is back at the head position due to the change of the schedule. Afterwards, the algorithms waits until the submission of another job or until a machine becomes available. In the latter case, it starts the job at the head of the list as soon as the current time has reached or passed the required delay. 

Structurally, Algorithm~\ref{alg:online} is very similar to delayed-SPT on a single machine. It belongs to a common class of scheduling algorithms that only use the schedule generated so far and the parameters of a single job to schedule this job. All deterministic online algorithms with immediate commitment obviously belong to this class. In addition, this class includes some deterministic scheduling algorithms, like, for instance, SPT, Largest Processing Time first, and Earliest Due Date first, which require an additional sorting of all jobs before schedule generation. To apply this type of algorithm to online scheduling with delayed commitment, we must at least repeat the sorting step after each submission of a new job.

\subsection{Two Machine Environment}
\label{sec:two_machines}

We focus our analysis on the two machine environment and show that Algorithm~\ref{alg:online} produces the optimal competitive ratio $\mathcal{R}_2$ for this environment. Since the proof is elaborate, we split it into several parts. First we introduce some relations regarding the scheduling of a new job. If both machines are available then we simply delay the first job $J_j$ in the SPT-list until $t_1\cdot p_j$. Therefore, we consider a scenario with one machine being busy until time $m_1$ and $t_1 \cdot p_j<m_1$ for the currently first job $J_j$ in the SPT-list such that the scheduling of job $J_j$ leads to the availability $m_2=\alpha \cdot m_1$ of the second machine. We describe the parameters of Eq.~\eqref{eq:calc_release} and ~\eqref{eq:calc_solution} as functions of $m_1$ and $\alpha$: 
\begin{align*}
\frac{\sum_{i=1}^2 m_i}{2} & = \frac{m_1}{2}\cdot (1+\alpha) \\
\left(\frac{\sum_{i=1}^2 m_i}{2}\right) ^2 & =  \frac{m_1^2}{4}\cdot (1+\alpha)^2 \\
\frac{\sum_{i=1}^2 m_i^2}{2} & = \frac{m_1^2}{2}\cdot \left( 1+\alpha^2\right) 
\end{align*}
Then Eq.~\eqref{eq:calc_solution} yields the earliest start time $r(\alpha)$ of job $J_j$:
\begin{align}
\label{eq:start_time}
r(\alpha) & = \frac{m_1}{2\mathcal{R}_2}\cdot \left(1+\alpha - \left|(1-\alpha)\right|\cdot \sqrt{\mathcal{R}_2-1} \right)
\end{align}
Equation~\eqref{eq:start_time} is only valid for $r(\alpha)< m_1$ since otherwise, there is no interval with both machines being busy and we start job $J_j$ at time $t_1\cdot p_j$. If a job $J_j$ completes at time $C_j=\alpha \cdot m_1$ and observes $\mathcal{R}_2$ then it has at most processing time
\begin{align}
\label{eq:processing_time2}
p_j(\alpha) & = \alpha\cdot m_1-r(\alpha) = m_1 \cdot \left(\alpha - \frac{1}{2\mathcal{R}_2}\cdot \left(1+\alpha - \left|(1-\alpha)\right|\cdot \sqrt{\mathcal{R}_2-1} \right)\right).
\end{align}   
For a given value $\alpha$, we can eliminate $\alpha$ and determine the start time in dependence of $p_j$.
\begin{align}
\label{eq:small_alpha}
r(p_j) & = \frac{1+\sqrt{\mathcal{R}_2-1}}{2\mathcal{R}_2-1-\sqrt{\mathcal{R}_2-1}}\cdot p_j + \frac{m_1\cdot (1-\sqrt{\mathcal{R}_2-1})}{2\mathcal{R}_2-1-\sqrt{\mathcal{R}_2-1}} = 1.28508 \cdot p_j + 0.19288\cdot m_1 & \mbox{   for   } \alpha \leq 1 \\
\label{eq:large_alpha}
r(p_j) & = \frac{1-\sqrt{\mathcal{R}_2-1}}{2\mathcal{R}_2-1+\sqrt{\mathcal{R}_2-1}}\cdot p_j + \frac{m_1\cdot (1+\sqrt{\mathcal{R}_2-1})}{2\mathcal{R}_2-1+\sqrt{\mathcal{R}_2-1}}= 0.09219 \cdot p_j + 0.61423 \cdot m_1 & \mbox{   for   } \alpha > 1
\end{align}
Equations~\eqref{eq:small_alpha} and \eqref{eq:large_alpha} provide formal expressions for the content of Lemma~\ref{lem:release_date}.

Note that Eq.~\eqref{eq:calc_release} yields
\begin{align}
\label{eq:calc_t1}
\frac{1-\sqrt{\mathcal{R}_2-1}}{2\mathcal{R}_2-1-\sqrt{\mathcal{R}_2-1}} & = \frac{t_1}{1+t_1}.
\end{align}

Our main theorem states the tight performance of the Algorithm~\ref{alg:online} for the two-machine environment.
\begin{theorem}
\label{thm:cr2}
Algorithm~\ref{alg:online} guarantees competitive ratio $\mathcal{R}_2$ for the  minimization of the total completion time with delayed commitment on two parallel identical machines. 
\end{theorem}
For the proof, we use induction in the number of inversions. For a better understanding, we formulate the steps of the proof with the help of additional lemmas. The next lemma establishes the induction base.

\begin{lemma}
\label{lem:scenario1}
If Algorithm~\ref{alg:online} has produced an SPT-schedule $S$ on two parallel identical machines then the ratio of the total completion time of $S$ over the total completion time of an optimal schedule never exceeds $\mathcal{R}_2$.   
\end{lemma}

\begin{proof}
As our first scenario, we assume that all jobs have the same processing time $p=1$ and submission time $0$. Further we use the two delays $t_1=0.23898$ and $t_2=0.85321$, see Table~\ref{tab:ti}. The claim holds for the first two jobs, see Lemma~\ref{lem:unique_delay}. Due to  
$$0.09219 + 0.61423 \cdot (1+t_2) = 1.23049 < 1.23898 = 1+t_1, $$
Algorithm~\ref{alg:online} starts the third job immediately after the completion of the first job on the same machine as the first job. Similarly, it starts the forth job immediately after the completion of the second job on the same machine as the second job due to 
$$0.09219 + 0.61423 \cdot (2+t_1) = 1.46744 < 1.85321 = 1+t_2. $$
Due to Lemma~\ref{lem:release_date}, there is no intermediate idle time in the SPT-schedule for all other jobs. Therefore, the completion time of every job allocated to the first machine in the online schedule has delay $t_1$ compared to the optimal completion time of this job. For jobs allocated to the second machine, the corresponding delay is $t_2$. Therefore, the claim holds if all jobs have the same processing times. 

Due to Eq.\eqref{eq:large_alpha}, an increase of the processing time from $p$ to $p'$ for two subsequent jobs produces for the first job a change of the delay and the completion time which are upper bounded by
\begin{align*}
 r(p')-r(p) & = 0.09219 \cdot (p'-p) < t_1\cdot (p'-p) \\
 p'+r(p')-p-r(p) & =  1.09219 \cdot (p'-p), 
\end{align*}
and for the second job a change of the delay with the upper bound
\begin{align*}
 r(p')-r(p) & = 0.09219 \cdot (p'-p) + 0.61423 \cdot (m_1'-m_1) = 0.09219 \cdot (p'-p) + 0.61423 \cdot 1.09219 \cdot (p'-p) \\
 & = 0.76305 \cdot (p'-p) < t_2\cdot (p'-p). 
\end{align*}
Since the upper bounds of the delays of the first and the second jobs are upper bounded by factors $t_1$ and $t_2$, respectively, the claim continues to hold if the jobs have non-decreasing processing times and the common submission time $0$. 

Finally, we consider different submission times. These submission times have no impact in an SPT-schedule if Algorithm~\ref{alg:online} does not start any job at its submission time directly following some idle time in the schedule. Then the completion time ratio cannot increase. If for any two jobs $J_j$ and $J_k$ with $p_j<p_k$, we have $r_j \leq r_k$ then the claim also holds since any additional delay of a submission time cannot produce a larger delay for any following job in the SPT-schedule. Therefore, we assume that job $J_j$ starts at its submission time and another job $J_k$ with $p_k>p_j$ has submission time $r_k<r_j$. Since Algorithm~\ref{alg:online} does not start job $J_k$ before $r_j$, we have $r(p_k)\geq r_j$. Instead of job $J_j$, we consider job $J_j'$ such that $J_j'$ completes at the same time as $J_j$ while we have $r_j'<r(p_j')\leq r_j$ and $p_j<p_j'\leq p_k$. The replacement of job $J_j$ by job $J_j'$ does not change the total completion time of the SPT-schedule while it cannot increase the optimal total completion time. Since the claim holds for the schedule with replacement, it also holds for the original schedule.
\end{proof}

For the induction step, we first consider an inversion that produces an additional delay for the job with the smaller processing time. Using Lemma~\ref{lem:release_date} and Eq.~\eqref{eq:small_alpha} and \eqref{eq:large_alpha}, we determine that the worst case for this inversion occurs if there is a single long job before the inversion.  
\begin{lemma}
\label{lem:singlejob2}
Consider an environment with two parallel identical machines and assume that long job $J_1$ with $p_1=1$ and $r_1=0$ has started as the first job on machine $1$ at time $t_1=0.23898$. Immediately after starting job $J_1$, there is a submission of several jobs with submission time $t_1$. The processing time of each such job is less than $p_1$. Then the competitive ratio of Algorithm~\ref{alg:online} does not exceed $\mathcal{R}_2$.   
\end{lemma}

\begin{proof}
Job $J_1$ and the next job $J_2$ complete at $m_1=t_1+1$ on machine $1$ and $\alpha \cdot m_1= \alpha \cdot (t_1+1)$ on machine $2$, respectively, see Eq.~\eqref{eq:start_time}. Due to $p_2<p_1$, $t_1< \alpha\cdot m_1 < t_2+p_1$ holds. Job $J_2$ starts with the additional delay $\delta (p_2) = r(p_2)-t_1$. 

We consider different scenarios. First, we assume that $p_2\leq p(\alpha=1)=0.43762$ and Algorithm~\ref{alg:online} only allocates job $J_1$ to machine 1. Due to Eq.~\eqref{eq:small_alpha} and \eqref{eq:large_alpha}, there is no intermediate idle time between any two jobs on machine $2$ in the online schedule. Since all jobs except job $J_1$ have the same submission time, we obtain the largest ratio if all jobs with submission time $t_1$ have the same processing time $p_2$, that is, the total processing time of these jobs is minimal. If there is a total of $k$ jobs then the total completion time of the online schedule $S$ is
$$ \sum C_j(S) = t_1+1+ (k-1) \cdot (t_1 + 1.28508\cdot p_2) + \frac{(k-1)\cdot k}{2} \cdot p_2. $$
If job $J_1$ starts at time $0$ in the optimal schedule then the total completion time of the optimal schedule is
$$\sum C_j^* = 1 + (k-1) \cdot t_1 + \frac{(k-1)\cdot k}{2} \cdot p_2. $$ 
For any given value of $k$, the largest possible processing time $p_2$ produces the largest ratio. More precisely, we have $p_2=p(1)$ for $k\leq 3$. The resulting ratios for $k=2$ and $k=3$ are $1.47797 <\mathcal{R}_2$ and $1.48865 < \mathcal{R}_2$, respectively. For $k\geq 4$, the requirement of allocating all jobs except $J_1$ to the same machine in the optimal schedule limits processing time $p_2$. All values of $k\geq 4$ are not relevant due to the small makespan on machine 2 in both schedules, for instance, for $k=4$ and $k=5$, we obtain $p_2=t_1$ with ratio $1.36826$ and $p_2=\frac{t_1}{2}$ with ratio $1.27079$, respectively. The ratio cannot increase if a very small $p_2$ causes an intermediate idle time on machine $2$ since the online total completion time increases by $t_1+2.28508p_2$ while the optimal total completion time increases by at least $t_1+2p_2$. 

Adding another job with processing time $p_2=p(1)$ requires allocation of this job to machine $1$, moves job $J_1$ to the last position in the optimal schedule, and produces ratio $1.40713<\mathcal{R}_2$. Any further addition of a job with processing time $p_2$ results in an online completion time of this job  that is at most $1-p_2$ larger than its optimal completion time. Due to its minimal optimal completion time $t_1+2p_2$, the ratio is at most $\frac{1+t_1+p_2}{t_1+2p_2}=1.50473<\mathcal{R}_2$. Therefore, the claim holds for this scenario.  

If job $J_1$ starts last in the optimal schedule then the total completion time of the optimal schedule is 
\begin{align*}
\sum C_j^* & = k \cdot t_1 + \frac{k^2+2k-4}{4} \cdot p_2 + 1 & k \mbox{ even}, \\ 
\sum C_j^* & = k \cdot t_1 + \frac{k^2+2k-3}{4} \cdot p_2 + 1 & k \mbox{ odd}.
\end{align*}    
We always obtain a smaller ratio for even values of $k$ than for odd values of $k$ and the ratio is largest for the largest value $k$, that is, we come close to our key pattern with the exception of the additional delay, see Eq.~\eqref{eq:small_alpha}, and the impact of job $J_1$. For $k\rightarrow \infty$, both effects increase the total completion time of the optimal schedule and the total completion time of the online schedule by $t_1+ 0.5 +1$ and $t_1+1+1.28508 < \mathcal{R}_2 \cdot (t_1+1.5)$, respectively. Therefore, the claim also holds for this scenario even if we add additional jobs with processing time $p_2$.    

Next we address $\alpha > 1$, that is $0.43762 = p(1)< p_2 < 1$. Then Algorithm~\ref{alg:online} alternatively allocates the jobs to both machines such that the online schedule executes $k_1$ and $k_2 \in \{k_1-1,k_1\}$ jobs on machines $1$ and $2$, respectively. First we assume no intermediate idle time between any pair of jobs allocated to the same machine and again obtain the largest ratio if all jobs have the same processing time. Using Eq.~\eqref{eq:large_alpha} and $0.61423\cdot (t_1+1) -t_1=0.52203$, the total completion time of the online schedule $S$ is
\begin{align*}
\sum C_j(S) & = k_1\cdot (t_1+1) + \frac{(k_1-1)\cdot k_1}{2}\cdot p_2 + k_2\cdot (t_1+0.09219 \cdot p_2+0.52203) + \frac{k_2\cdot (k_2+1)}{2}\cdot p_2.
\end{align*}
If job $J_1$ starts last in the optimal schedule then there is an SPT-schedule that  allocates each job to the same machine as in the online schedule. The resulting total completion time is
\begin{align*}
\sum C_j^* & = \frac{(k_1-1)\cdot k_1}{2}\cdot p_2 + k_1\cdot t_1+ (k_1-1)\cdot p_2 + 1 +\frac{k_2\cdot (k_2+1)}{2}\cdot p_2+k_2\cdot t_1.
\end{align*}
If job $J_1$ starts at time 0 in the optimal schedule then there is an optimal schedule that allocates each job to the same machine as in the online schedule with the  exception of the last job if $p_2< 1-t_1$ and $k_1=k_2$ hold. 
\begin{align*}
\sum C_j^* & = k_1 + \frac{(k_1-1)\cdot k_1}{2}\cdot p_2 + k_2\cdot t_1+ \frac{k_2\cdot (k_2+1)}{2}\cdot p_2 & \mbox{identical allocation} \\
\sum C_j^* & = k_1-1 \frac{(k_1-2)\cdot (k_1-1)}{2}\cdot p_2 + (k_2+1)\cdot t_1+ \frac{(k_2+1)\cdot (k_2+2)}{2}\cdot p_2 & \mbox{swapping of the last job} 
\end{align*}
In those situations, the maximization of the ratio $R$ requires $p_2(\alpha)$ to be as small as possible for all values of $k_1$ and $k_2$, that is, $p_2= p(1)=0.43762$. Then we already know that the claim holds. 

Finally, it is possible that a job may produce another delay if the job has a larger processing time than $p(1)$. Remember that we already know that the claim holds if all jobs except job $J_1$ have the same processing time and this processing time is at least $p(1)$. Then we apply the same approach as in the proof of Lemma~\ref{lem:scenario1} to show that the increase of the processing time to $p'$ does not increase the ratio. Therefore, the claim also holds in this scenario.
\end{proof}

While the first small job in the scenario of Lemma~\ref{lem:singlejob2} results in an additional delay time, such delay does not occur for the next small job with the same or less processing time than the first small job since the delay only depends on the availability of the machine with more load and the processing time of the job, see Eq.~\eqref{eq:small_alpha}. Therefore, we must also address small jobs without any additional delay but a non-negligible processing time. The next lemma considers the impact of such small jobs without any delay time. We compare a pattern of small jobs with a key pattern that uses the same machine availability and has the same total processing time. Contrary to the key pattern, the small job pattern may not produce the same makespan on both machines. We use $\mathcal{R}(\delta)$ to denote the total completion time ratio of the key pattern since the processing time of the jobs in the key pattern has the arbitrarily small value $\delta$.  

\begin{lemma}
\label{lem:r_delta}
Assume machine availability $m_1$ and $m_2\geq m_1$ and a sequence of jobs in SPT-order  in a restricted schedule $S$ without any starting idle time or intermediate idle time.
The ratio of the total completion time of this sequence in the restricted schedule over the total completion time in the optimal schedule of this sequence is upper bounded by the maximum of the total completion time ratio of the corresponding key pattern and $1.25$. 
\end{lemma}

\begin{proof}
Since the restricted schedule and the optimal schedule for the small jobs are SPT-schedules, we obtain the largest total completion time ratio if all small jobs have the same processing time. Let us assume $k$ such small jobs with processing time $p$. We define $\kappa = \min\{k,\frac{m_2-m_1}{p}\}$. For the corresponding key pattern, we apply Eq.~\eqref{eq:delta_ratio} and obtain its total completion ratio
\begin{align*}
\mathcal{R}(\delta) & = \frac{\frac{\left(k-\kappa\right)^2}{4}\cdot p + (k-\kappa)\cdot m_2+\frac{\kappa^2}{2}\cdot p+\kappa\cdot m_1}{\frac{k^2}{4}\cdot p + k\cdot r} \leq \mathcal{R}_2
\end{align*}

For the system of small jobs, there are two expressions for the optimal total completion time:
\begin{align*}
\sum C_j^* & = \frac{k^2}{4}\cdot p + \frac{k}{2}\cdot p + k\cdot r &  k \mbox{ even} \\
\sum C_j^* & = \frac{k^2}{4}\cdot p + \frac{k}{2}\cdot p + \frac{1}{4}\cdot p + k\cdot r &  k \mbox{ odd} \\
\end{align*}

For $\kappa=k$, the total completion time of the restricted schedule $S_r$ is 
\begin{align*}
\sum C_j (S_r) & = \frac{k^2}{2}\cdot p +\frac{k}{2}\cdot p + k\cdot m_1.  
\end{align*}

IF $\kappa>k$ is an integer, then there are two expressions for the total completion time of the restricted schedule $S_r$: 
\begin{align}
\nonumber \sum C_j (S_r) & = \frac{(k-\kappa)^2}{4}\cdot p + \frac{\kappa^2}{2}\cdot p + \frac{k}{2}\cdot p + (k-\kappa) \cdot m_2 + \kappa \cdot m_2 & k-\kappa \mbox{ even} \\
\label{eq:odd_value}
\sum C_j (S_r) & = \frac{(k-\kappa)^2}{4}\cdot p + \frac{\kappa^2}{2}\cdot p +\frac{k}{2}\cdot p + \frac{1}{4}\cdot p + (k-\kappa)\cdot m_2 + \kappa\cdot m_1 &  k-\kappa \mbox{ odd} 
\end{align}
If $\kappa>k$ is not an integer then Eq.~\eqref{eq:odd_value} is an upper bound for the total completion time of the restricted schedule $S_r$. 

We use $\mathcal{R}(p)$ to denote the total completion time ratio of the small jobs. A simple comparison yields $\mathcal{R}(p)< \mathcal{R}(\delta)\leq \mathcal{R}_2$ unless $k$ is even and $\kappa$ is odd. In letter case, we obtain the ratio  
\begin{align*}
\mathcal{R}(p) & = \frac{\frac{(k-\kappa)^2}{4}\cdot p + \frac{k}{2}\cdot p +\frac{1}{4}\cdot p + \frac{\kappa^2}{2}\cdot p + (k-\kappa)\cdot m_2 +\kappa\cdot m_1}{\frac{k^2}{4}\cdot p + k\cdot r+\frac{k}{2}\cdot p}.
\end{align*}
Therefore, the claim holds due to $\frac{k+\frac{1}{2}}{k}\leq 1.25$ for $k\geq 2$.
\end{proof}

After having collected the main ingredients, we can now describe the proof of Theorem~\ref{thm:cr2}.

\begin{proof}
We reduce the submission time $r_j$ of a job $J_j$ to the smallest time that does not change the online schedule. Such reduction cannot decrease the total completion time ratio since it does not increase the total completion time of the optimal schedule.

We use induction in the number of inversions. Lemma~\ref{lem:scenario1} describes the induction base. The proof of Lemma~\ref{lem:scenario1} also shows that any SPT-schedule generated by  Algorithm~\ref{alg:online} does not include a job $J_j$ with a total delay of more than $t_2\cdot p_j$ compared to the start time of this job in the optimal schedule. 

We consider two types of inversions and start with an inversion that results in an additional idle time for the first job of this inversion. Lemma~\ref{lem:singlejob2} addresses the worst case with the smallest value of availability $m_1$ in relation to the processing time of the first job and shows that ratio $\mathcal{R}_2$ remains valid for this scenario. 

The other type of inversion produces no additional idle time for the first job of the inversion. Such situation occurs if the common submission time $r_{k}$ of the jobs of inversion $k$ is already past the start time of  Eq.~\eqref{eq:small_alpha} or \eqref{eq:large_alpha} or if the completion times of the running jobs on both machines are larger than $r_{k}$. Due to Lemma~\ref{lem:r_delta}, Algorithm~\ref{alg:online} only generates schedules such that the completion time ratio of the jobs of inversion $k$ is less than $\mathcal{R}_2$ if we ignore all other jobs in the optimal schedule. 

Finally, we combine the jobs of a new inversion $k$ with all previous or \textit{old} jobs. There is no impact on the total completion time ratio if an old job starts before any job of inversion $k$ both in the online schedule and in the optimal schedule. 

If an old job starts before any job of inversion $k$ in the online schedule but not in the optimal schedule then the total completion time ratio cannot increase. 

We must address the remaining two cases more carefully. Assume that an old job starts after one or more jobs of inversion $k$ both in the online schedule and in the optimal schedule. Since there is no intermediate idle time between the jobs of inversion $k$ in the online schedule, the total completion time ratio cannot increase if neither the old job nor the first job of the inversion incurs any additional delay. The worst case of such an additional delay occurs if both machines are available at the same time, there are two old jobs with the same processing time and the machine availability is as small as possible with respect to this processing time of the old jobs. Due to the proof of Lemma~\ref{lem:scenario1}, we can assume that those two old jobs have the same processing time $p$ as the preceding old job, that is, both machines are available at time $(t_1+1)\cdot p$. The first of both old jobs starts at this time without any additional delay due to $t_1+1>t_1$. For the following old job, we have $m_1 = (t_1+2) \cdot p$ resulting in start time $r = 0.09219\cdot p + 0.61423 \cdot (t_1+2)\cdot p = 1.46744\cdot p$, see Eq.~\eqref{eq:large_alpha}, and an additional idle time $r-(t_1+1)\cdot p= 0.22846\cdot p < t_1\cdot p$. Even if we add the additional idle time of the first job of inversion $k$ then we obtain $0.22846 \cdot p + 1.28508 \cdot 0.43762\cdot p = 0.79084\cdot p < t_2\cdot p$ in the worst case if inversion $k$ only comprises a single job with submission time $t_1\cdot p$ and processing time $0.43762 \cdot p$, see the proof of Lemma~\ref{lem:singlejob2}. Due to the limits of such upper bounds, such situation cannot increase the total completion time ratio beyond $\mathcal{R}_2$. For instance, the completion time ratio for this job sequence is $\frac{3t_1+5+1.46744}{2t_1+4+2\cdot 0.43762}= 1.34207<\mathcal{R}_2$. 

Finally, we consider old jobs that do not start after the first job of inversion $k$ in the optimal schedule while they start after the jobs of inversion $k$ in the online schedule. We temporarily remove all old jobs that start after the jobs of inversion $k$ in both schedules. Let $p_{old}$ be the processing time of the last old job in the online schedule before any job of inversion $k$. Then $i$ is the smallest integer such that $i\cdot p_{old}$ is not smaller than the total processing time of all jobs of inversion $k$. We remove inversion $k$ by replacing all jobs of inversion $k$ with $i$ jobs with processing time $p_{old}$ in the online schedule. Due to the induction hypothesis, the total completion time ratio does not exceed $\mathcal{R}_2$. Next we reverse the replacement of the jobs of inversion $k$. This reversal involves some processing time reduction.  Reversing the replacement of the job of inversion $k$ does not change the total completion of the old jobs in the optimal schedule while it does not increase the total completion time of the old jobs in the online schedule. Due to Lemma~\ref{lem:r_delta}, it does not increase the contribution of the jobs of inversion $k$ beyond $\mathcal{R}_2$.  Now we attach all old jobs that start after the jobs of inversion $k$ after the other old jobs in the online schedule. Due to Lemma~\ref{lem:scenario1}, this addition does no increase the total completion time ratio. Rearranging the old jobs in SPT-order does not increase the total completion time of the online schedule while the total completion time of the optimal schedule remains unchanged.
\end{proof}

\subsection{Environments with More Machines}
\label{sec:more_machines}

We discuss the application of Algorithm~\ref{alg:online} to parallel identical machine environments with more than 2 machines. The competitive ratio $\mathcal{R}_2$ also holds for all $m$-machine environments if $m$ is even. 

\begin{theorem}
\label{thm:even_number}
Algorithm~\ref{alg:online} guarantees competitive ratio $\mathcal{R}_2$ for the  minimization of the total completion time with delayed commitment on $m$ parallel identical machines if $m$ is even.
\end{theorem}

\begin{proof}
We partition the machines into two groups: machines $1$ to $k=\frac{m}{2}$ and machines $k+1$ to machine $m$. Then we relate the first group to the first machine and the second group to the other machine in the two-machine environment.

Assume that a machine in the two-machine environment completes job $J_j$ at time $C_j(S)$ in the online schedule. If all machines of the corresponding group complete their jobs $J_{j,1}$ to $J_{j,k}$ at the same time $C_j(S)$ in their online schedule then Algorithm~\ref{alg:online} with target ratio $\mathcal{R}_2$ starts job $J_{j,k}$ at time $C_j(S)-p_j$. The start times of all other jobs are earlier since the small jobs of the key pattern have more resources. Therefore, the total processing times of jobs $J_{j,1}$ to $J_{j,k}$ is larger than $m\cdot p_j$. 

If not all jobs $J_{j,1}$ to $J_{j,k}$ complete at the same time then we determine the average completion time of these jobs and use a job $J_j$ that completes at this average completion time on the corresponding machine in the two-machine environment. Such constellation increase the total processing time of jobs $J_{j,1}$ to $J_{j,k}$ even more than the constellation with a common completion time since some small jobs of the key pattern can start earlier, see the proof of Lemma~\ref{lem:trend}. Since the total processing time increases in each comparison, jobs $J_{j,1}$ to $J_{j,k}$ complete later than job $J_j$ in the corresponding optimal schedules and the competitive ratio of the $m$-machine environment is less than $\mathcal{R}_2$.
\end{proof}

We can generalize Theorem~\ref{thm:even_number}: if we know a competitive ratio for any other number $m$, then this competitive ratio is an upper bound of the competitive ratio for a machine environment that comprises a multiple of $m$ machines. Note that the gap between upper and lower bounds increases with growing machine numbers.

To reduce this gap, we need a better analysis of environments with more than 2 machines. To this end, we must adapt some lemmas of Section~\ref{sec:two_machines}. The total lack of any intermediate idle time for any long job after the first $m$ jobs in a scenario with a common processing time and a common submission time for all jobs is a specific property of $m=2$, see Lemma~\ref{lem:scenario1}. For $m>2$, some jobs among the second generation of long jobs (jobs $J_{m+1},\ldots, J_{2m}$) do not start immediately after the completion of the previous long job on the same machine, see Table~\ref{tab:2delay}. The reason for this property is the immediate or almost immediate start of some long jobs in the solutions of Section~\ref{sec:lower_bounds}. Since the additional delays of these long jobs are rather small and never exceed $\mathcal{R}_m$, we can extend Lemma~\ref{lem:scenario1} to all other machine numbers as well. 

\begin{table}[ht]
\begin{center}
\begin{tabular}{|c|c|c|c|c|c|c|c|c|c|c|}
\hline
$m$ & 3 & 4 & 5 & 6 & 7 & 8 & 9 & 10 & 100 & 1000 \\ \hline
Jobs & 4 & 5 & 6-7 & 8 & 9 & 10-11 & 11-12 & 12-13 & 120-134 & 1199 - 1337 \\
$\Delta_{max}$ & 0.10260 & 0.05350 & 0.00426 & 0.05137 & 0.08227 & 0.04766 & 0.03560 & 0.06178 & 0.06270 & 0.06120 \\
\hline
\end{tabular}
\end{center}
\caption{\label{tab:2delay} Long jobs of the second generation with a delay and the maximum additional delay $\Delta_{max}$ of these jobs for different numbers $m$ of machines}
\end{table}

Similarly, we can extend Lemma~\ref{lem:r_delta} to all other machine numbers.

An extension of Eq.~\eqref{eq:small_alpha} and \eqref{eq:large_alpha} is conceptually possible. However, the delay depends on the processing time and $m-1$ variables describing the machine availability. Moreover, we must consider more than two cases for $\alpha$. Therefore, we were not able to find explicit descriptions of the relations in general. 

Finally, there is no equivalent lemma for $m>2$ to Lemma~\ref{lem:singlejob2}. On the contrary, we can show that Algorithm~\ref{alg:online} cannot achieve competitive ratio $\mathcal{R}_3$ for a three-machine environment.  
\begin{theorem}
\label{thm:tightness}
The competitive ratio of Algorithm~\ref{alg:online} is larger than $\mathcal{R}_3$ in a three-machine environment.
\end{theorem}

\begin{proof}
Due to Theorem~\ref{thm:lbu} and Table~\ref{tab:ti}, an optimal online algorithm must start the first of three identical long jobs $J_1$ with processing time $p_1=1$ at time $t_1=0.02991$ to achieve competitive ratio $\mathcal{R}_3$. Assume that immediately after the start of this job, there is a submission of two identical jobs $J_2$ with $p_2=0.58679$ and $r_2=t_1$. Algorithm~\ref{alg:online} uses Eq.~\eqref{eq:calc_release} with $\mathcal{R}_3$ to determine the start times $0.44312$ and $0.7081$ for these jobs. The resulting ratio of the total online completion time over the total optimal completion time is less than $\mathcal{R}_3$. 

If there is only a single submission of a job $J_1$ then Algorithm~\ref{alg:online}  produces the same starting times for this job and the two jobs $J_2$. The ratio of the total online completion time over the total optimal completion time for these three jobs is 
\begin{align*}
\frac{2\cdot 1.02991+0.7081+0.58679}{1+2\cdot 0.02991 + 2 \cdot 0.58679} & = 1.50207 > \mathcal{R}_3.
\end{align*} 

However, if we select an increased start time $0.10107>t_1$ for the first job then this pattern produces a ratio of the total online completion time over the total optimal completion that does not exceed $\mathcal{R}_3$ even in the worst case with processing time $p_2= 0.62756$ and start times $0.47391$ and $0.7573$. Therefore, we can only find an optimal delay for the first job $J_1$ if we consider whether and which other long jobs have already been submitted.
\end{proof}
The job example used in the proof of Theorem~\ref{thm:tightness} is related scenario describe in Lemma~\ref{lem:singlejob2}. In the same fashion, we can construct corresponding examples for $m>3$ and extend the validity of Theorem~\ref{thm:tightness} to all $m\geq 3$. 

Note that Theorem~\ref{thm:tightness} does not establish a larger lower bound than Theorem~\ref{thm:lbu}. To settle the problem of a tight competitive ratio, we must either find a new lower bound construction or use an algorithm that considers more jobs in the SPT-list.

\section{Total Weighted Completion Time}
\label{sec:weights}

In the single machine environment, we use the corresponding delayed-WSPT (Weighted Shortest Processing Time first) algorithm to achieve the same competitive ratio as for the objective without weights. Sitters~\cite{Sit10} presented an algorithm that achieves the same competitive ratio for the total weighted completion time objective and the total completion time objective. This result suggests the conjecture that the optimal competitive ratios for total completion time scheduling with and without weights are the same for any parallel identical machine environment. In the next theorem, we disprove this conjecture and show that no algorithm can produce  competitive ratio $\mathcal{R}_2$ in an environment with two machines for the total weighted completion time.

\begin{theorem}
\label{thm:lbuw2}
Any deterministic online algorithm cannot achieve the competitive ratio $\mathcal{R}_2$ for the minimization of the total weighted completion time with delayed commitment on two parallel identical machines. 
\end{theorem}

\begin{proof}
The analysis of Section~\ref{sec:two_machines} requires a balance between the delays of different jobs. Since we do not know the weights of jobs with a later submission time, we can overemphasize the contribution of those jobs by selecting large weights. This way, we generate an example that exceeds the lower bound ratio $\mathcal{R}_2$.

The adversary submits two identical jobs $J_1$ with processing time $p_1=1$, weight $w_1=1$, and submission time $r_1=0$. The online algorithm starts the first of these jobs at time $t_1$. Since this job pattern is identical to the starting job pattern in the proof of Theorem~\ref{thm:lbu}, we know from this proof that the algorithm must select delay $t_1$ to avoid a larger competitive ratio than $\mathcal{R}_2$.

The adversary submits at time $t_1$ another job $J_2$ with processing time $p_2=0.3$ and a sufficiently large weight $w_2$. The algorithm starts this job at time $t'$ and produces the competitive ratio
\begin{align}
\label{eq:ratio_simple}
\mathcal{R}(t') & = \frac{t_1+1+w_2\cdot (t'+p_2)+ t'+1+p_2}{1+w_2\cdot (t_1+p_2)+t_1+1+p_2} = \frac{2.3+t_1+t'+ w_2 \cdot (t'+0.3)}{2.3+t_1+ w_2\cdot (t_1+0.3)}.
\end{align} 
Since $w_2$ is sufficiently large, we need $t'\leq (\mathcal{R}_2-1)\cdot p_2 + \mathcal{R}_2 \cdot t_1= 0.53332$ for $\mathcal{R}(t')\leq \mathcal{R}_2$. Therefore, we must select $t'$ smaller than $0.53332$. 

Finally, the adversary submits many identical small jobs $J_3$ with processing time $p_3=0$ and a sufficiently high weight $w_3$ at time $t'$. The online schedule and the optimal schedule execute these jobs at $t'+p_2$ and $t'$, respectively. The competitive ratio approaches $1+\frac{p_2}{t'}$ and exceeds $\mathcal{R}_2$ for $t' < 0.54935$.
\end{proof}

Therefore, we have demonstrated that at least for the two-machine environment, the presence of weights leads to a larger competitive factor in total completion time scheduling. 

\bibliographystyle{ACM-Reference-Format}
\bibliography{schwiegelshohn}

\end{document}